\documentclass[twocolumn,superscriptaddress,showpacs,preprintnumbers,amsmath,amssymb]{revtex4}

\usepackage{graphicx}
\usepackage{float}
\usepackage[latin1]{inputenc}
\usepackage{natbib}
\usepackage{times}
\usepackage{bm}
\usepackage{amsfonts}
\usepackage{fancybox}
\usepackage{amssymb}
\usepackage[isu,scriptsize]{caption}

\def\urlprefix{}
\def\url#1{}


\begin{document}

\title{Impurity- and boundary-driven Collective to Kinetic transition in thermal transport}

\author{P. Torres}
\affiliation{Departament de F\'isica, Universitat Aut\`onoma de
Barcelona, 08193 Bellaterra, Catalonia, Spain}
\author{C. de Tomas}
\affiliation{Departament de F\'isica, Universitat Aut\`onoma de
Barcelona, 08193 Bellaterra, Catalonia, Spain}
\author{A. Lopeandia}
\affiliation{Departament de F\'isica, Universitat Aut\`onoma de
Barcelona, 08193 Bellaterra, Catalonia, Spain}
\author{X. Cartoix\`a}
\affiliation{Departament d'Enginyeria Electr\`onica, Universitat Aut\`onoma de
Barcelona, 08193 Bellaterra, Catalonia, Spain}
\author{F. X. Alvarez}
\affiliation{Departament de F\'isica, Universitat Aut\`onoma de
Barcelona, 08193 Bellaterra, Catalonia, Spain}
\email{Xavier.Alvarez@uab.es}

\date{\today}
\begin{abstract}
Several hitherto unexplained features of thermal conductivity in group IV materials, such as the change in the slope as a function of sample size for pure vs. alloyed samples and the fast decay in thermal conductivity for low impurity concentration, are described in terms of a transition from a collective to kinetic regime in phonon transport. We show that thermal transport in pure bulk silicon samples is mainly collective, and that impurity/alloy and boundary scattering are responsible for the destruction of this regime with an associated strong reduction in thermal conductivity, leaving kinetic transport as the only one allowed when those resistive scattering mechanisms are dominant. 
\end{abstract}
\maketitle

\section{Introduction}



It was already recognized by Peierls that normal scattering is the key point to understanding thermal conductivity~\cite{Peierls2001}. However, the preference for a kinetic point of view and the insufficient data on micro and nanoscale samples in those years led to the success of Callaway's proposal~\cite{Callaway1959}, and therefore the need to treat normal and resistive processes on a different footing was never properly acknowledged. It has only been in recent years, with the appearance of thermal transport measurements at the nanoscale~\cite{Li2003}, that the need for a deeper understanding has become evident, prompting some authors to work with modified expressions for the relaxation times, confinement effects or including new scattering mechanisms in the Callaway model \cite{Kazan2010,Mingo2003}.

Solutions of the Boltzmann equation with {\it ab initio} scattering rates with the proper inclusion of the role of normal (N) scattering have achieved in the last years a great level of accuracy, and the properties of large number of materials have been predicted without the need of any fitting parameter~\cite{Lindsay2013,Broido2007}. However, {\it ab initio} results are much less amenable to analysis and the computational cost may become unacceptable for samples with low dimensionality, showing alloy scattering, and/or strong surface effects. To overcome that, analytic expressions have been proposed from the {\it ab initio} calculations for the relaxation times in order to be used in simple approximate models~\cite{Ward2010}.

In recent papers a different approach, the Kinetic Collective Model (KCM), has been proposed~\cite{DeTomas2014,DeTomas2014a}. The model is based on a combination of a Kinetic and a Collective contribution, weighted by the relative importance of normal {\it vs.} resistive processes. The main differences with a pure kinetic model are the introduction of a hydrodynamic (collective) flow which has a unique value for the relaxation time for all the modes and uses a form factor to include boundary effects. 

Recent experimental works have demonstrated that pure kinetic models are not enough to understand thermal conductivity at short scales and times~\cite{Hoogeboom-Pot2015,Wilson2014}. In these works, collective flow has been used to explain the origin of the non-monotonous dependence of the thermal boundary resistance as a function of the size of the heater arising from ultrafast laser heating experiments ~\cite{Hoogeboom-Pot2015}. Also theoretically, collective transport has been successfully used to understand first-principles results on graphene thermal transport~\cite{Lee2015,Cepellotti2015},  showing that Poiseuille flow can be behind the high thermal conductivity of this material and thus converging with the KCM. Collective models will be necessary in next years in order to analise these experiments and simulations.

This Kinetic Collective Model has been able to predict thermal conductivity of natural and isotope rich silicon bulk, and micro and nanoscaled samples by fitting only the natural bulk sample~\cite{DeTomas2014}. In a following work~\cite{DeTomas2014a}, the calculations were extended to all Group IV materials, significatively extending the predictability range of previous analytical models. A noticeable point is that the expressions for the relaxation times that seem to work better with the KCM are those obtained in the works by Herring and Klemens~\cite{Herring1954,Klemens1955}, pointing to the possibility that the problems appearing in the last decades when predicting thermal conductivity may not be related to the relaxation times, but rather to the insuficient attention to this collective regime.


In this Letter we go a step forward in the KCM and predict the thermal conductivity in Si/Ge bulk and nanowire alloys from the {\em same} natural silicon fitting already used in Ref.~\onlinecite{DeTomas2014a}. From our predictions we demonstrate that the difficulty of predicting the thermal conductivity in this kind of samples is related to the change from a collective to a kinetic dominated flow. The fact that such a variety of materials, isotope fractions, alloy compositions, and length scales is correctly described by the KCM is a very strong indicator of its validity as an analysis, prediction and optimization tool.



From the KCM, the lattice thermal conductivity
\begin{equation}\label{kappa_final}
\kappa=\kappa_{\rm{k}}(1-\Sigma)+\kappa_{\rm{c}} F(L_{\rm eff}) \Sigma
\end{equation}
is obtained as the sum of the kinetic $\kappa_{\rm{k}}$ and the  collective $\kappa_{\rm{c}}$ contributions weighted by a switching factor
\begin{equation}\label{sigma}
 \Sigma=\frac{1}{(1+\left<\tau_N\right>/\left<\tau_R\right>)}\in [0,1]
\end{equation}
This factor determines the actual fraction of heat carried in each regime depending on the strength of normal ($\tau_N$) and resistive ($\tau_R$)  scattering times. Also, the form factor $F(L_{\mathrm{eff}})\in [0,1]$ modulates the hydrodynamic reduction of the flux on the collective term due to boundary effects, like in a Poiseuille flow, where $L_{\mathrm{eff}}$ is the effective size of the sample. Expressions for $\kappa_{\rm{k}}$ and $\kappa_{\rm{c}}$, $\Sigma$ and $F(L_{\mathrm{eff}})$ are given elsewhere~\cite{DeTomas2014,DeTomas2014a}.


In order to compute the properties of Si$_{1-x}$Ge$_{x}$ alloys, we need the phonon dispersion relations and the relaxation times. To obtain the dispersion relations for a stochiometry $x$, we perform lattice dynamics calculations within the Virtual Crystal Approximation (VCA) in an {\it ab initio} framework~\cite{Bellaiche2000}. Calculations were done with the {\sc Quantum ESPRESSO} package~\cite{Giannozzi2009}, which implements de Density Functional Theory (DFT)~\cite{Hohenberg1964,Kohn1965}, under the Local Density Approximation in the parametrization of Perdew and Zunger~\cite{Perdew1981}. Core electrons were accounted for with norm-conserving pseudopotentials of the von Barth-Car type~\cite{VonBarthU.andCar,DalCorso1993}, and plane waves were cut off at an energy of 60 Hartree. For each composition $x$, the lattice parameter was adjusted until the pressure was less than 0.1~kbar. Solution of the ensuing dynamical matrix provided the sought dispersion relations and derived quantities.

In this kind of alloys four different relaxation times should be considered; three of them are significatively contributing to the total resistivity, and they should be combined using Mathiessen's rule
\begin{equation}
\frac{1}{\tau_{R}}=\frac{1}{\tau_{I}}+\frac{1}{\tau_{B}}+\frac{1}{\tau_{U}}
\end{equation}
where $\tau_{B}$ is the boundary term, $\tau_{U}$ the umklapp contribution and $\tau_{I}$ the impurity term. The fourth element is the normal scattering term $\tau_{N}$. All these terms may depend on $x$ and the phonon dispersion. 

For boundary scattering we use 
\begin{equation}
\label{tauB}\tau_{B,x}=L_{\mathrm{eff}}/v_{x}
\end{equation}
where $L_{\mathrm{eff}}$ is the characteristic length of the sample and $v_{x}$ is the group velocity for stoichiometry $x$. 

For Umklapp and Normal scattering we use the same expressions as in \cite{DeTomas2014}:
\begin{equation}
\tau_{N,x}=\frac{1}{B'_{N,x}T}+\frac{1}{B_{N,x}T^{3}\omega_j^2[1-\exp(-3T/\Theta_D)]}\quad.
\label{tauN}
\end{equation}
and
\begin{equation}
\tau_{U,x}=\frac{1}{B'_{U,x}T}+\frac{\exp(\Theta_U/T)}{B_{U,x}\omega_j^4T[1-\exp(-3T/\Theta_D)]}\quad.
\label{tauN}
\end{equation}
where $\Theta_{D}$ is the Debye temperature, $\Theta_{U}$ the Umklapp freezing temperature , and $B_{U/N,x}$ the composition dependent parameters, which can be calculated from those of pure silicon $B_{USi}$ with~\cite{Morelli2002,Schlomann}
\begin{eqnarray}
\label{eq:BB}B_{(U/N)x}=f_{(U/N)x}B_{(U/N)Si}\\ 
\label{eq:BA}B'_{(U/N)x}=f'_{(U/N)x}B'_{(U/N)Si}
\end{eqnarray}
and using only material properties
\begin{equation}
\label{eq:fs}f_{x}= \frac{\left[V/M v^5\right]_{x}}{\left[V/M v^5\right]_{Si}}
\,\,
; \,\,\,f'_{x}= \frac{\left[1/M v V^{1/3}\right]_{x}}{\left[1/M v V^{1/3}\right]_{Si}}.
\end{equation}

Following this procedure, the only fitting values are those of pure Silicon: $B_{USi}=2.8\cdot 10^{-46}$ s$^{3}$K$^{-1}$ and $B_{NSi}=3.9\cdot 10^{-23}$ s$^{-1}$K$^{-1}$, $B'_{USi}=7\cdot 10^{8}$ s$^{-1}$K$^{-1}$ and $B'_{NSi}=7.5\cdot 10^{8}$ s$^{-1}$K$^{-1}$. The rest of the values are calculated from the dispersion relations. All the independent parameter values for the alloy can be computed from linear interpolation, except for the phonon velocity in Eqs.~(\ref{eq:fs}), which requires $v_{x}=((1-x)v^{-2}_{Si}+xv^{-2}_{Ge})^{-1/2}$.


\begin{figure}
\includegraphics[scale=0.7]{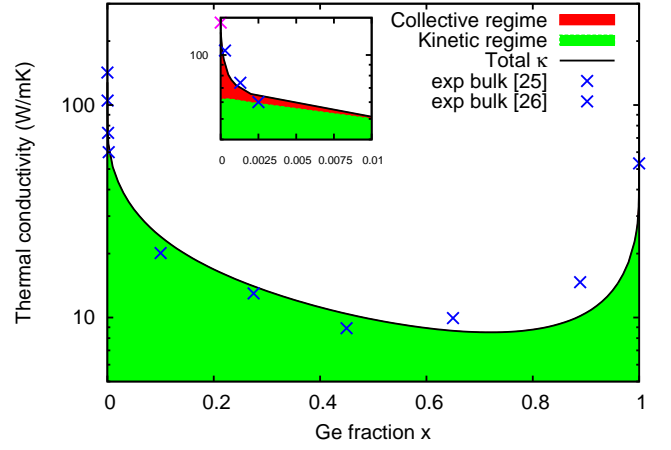}
\caption{Thermal conductivity in terms of Ge fraction $x$. Green filled line is the kinetic and red line the collective fraction of the thermal conductivity. The top of the green line is the total thermal conductivity. Crosses are the experimental thermal conductivity obtained from \cite{Abeles1963,Cahill2005}}\label{fig:alloy_bulk}
\end{figure} 

The alloy relaxation time needs a more detailed discussion. In single specie crystals the mass defect term describes the variability in isotopic abundance but in alloys it should also account for the variability in the force and lattice constants. Thus, including these three factors, the total impurity relaxation time can be expressed as~\cite{Klemens1955}:
\begin{equation}
\frac{1}{\tau_{I}}=\frac{\pi}{6}VS^{2} \omega^{2} D_{x}
\end{equation}
where as in Ref.~\onlinecite{Capinski1999}, instead of the Debye approximation, we use the DOS obtained from the full dispersion relations at stoichiometry $x$, $D_{x}$, and the variance term from Ref.~\onlinecite{Klemens1955}
\begin{equation}
\label{eq:S2}S^{2}=\frac{1}{12}\Gamma_{M}+\frac{1}{6}\Gamma_{v^{2}}+\frac{2Q^{2}\gamma^{2}}{3}\Gamma_{R} ,
\end{equation}
where 
\begin{equation}
\Gamma_{\alpha}=\sum_{i} x_{i}\left(\frac{\alpha_{i}-\bar{\alpha}_{x}}{\bar{\alpha}_{x}}\right)^{2} 
\end{equation}
is the coefficient of variance  of ($M$) mass, ($v^2$) squared velocity or ($R$) impurity radius, being  $\alpha_{i}$ the value for the isotope/species $i$ and $\bar{\alpha}_{x}$ the averaged value over all the atoms, $\gamma=1.7$ is the experimental Gr\"uneisen parameter and $Q=4$ is a factor depending on the geometry of the impurity (substitutional in this case). 

As expected, the second and third terms reduce to zero for pure silicon or germanium and the isotopic mass defect corresponding to the first term is the only one remaining.

Notice the simplicity in the relaxation times employed (\ref{tauB})-(\ref{eq:S2}), where only compositionally averaged values for the parameters are used. 

\begin{figure}
\includegraphics[width=\columnwidth]{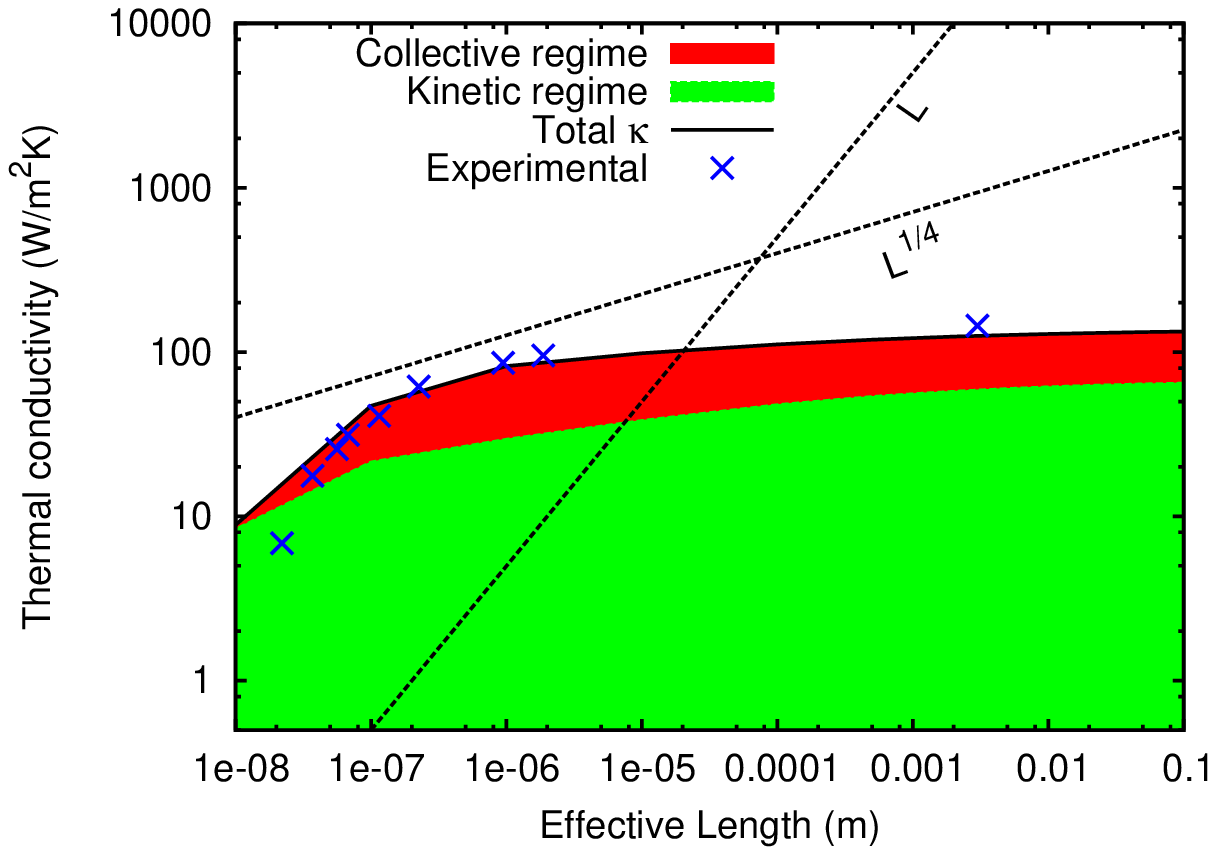}
\includegraphics[width=\columnwidth]{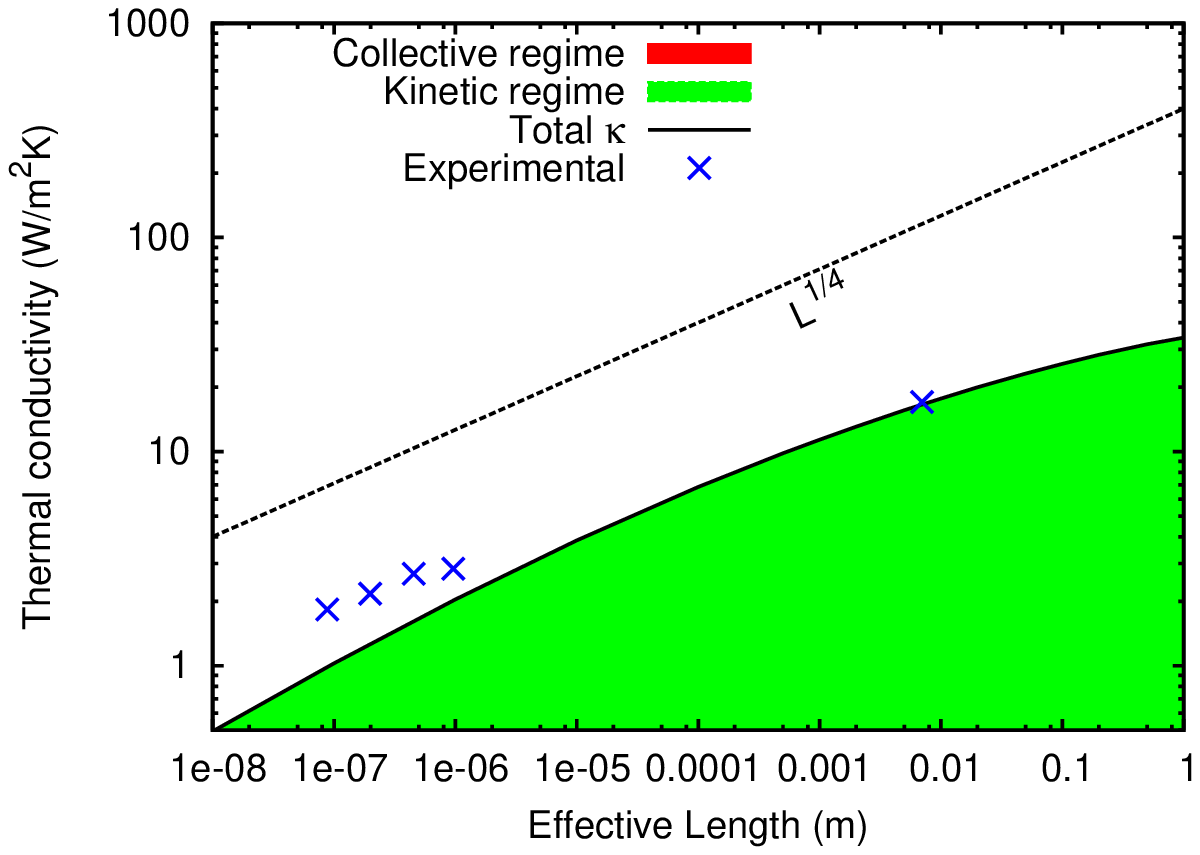}
\caption{Thermal conductivity in terms of size $L_{\mathrm{eff}}$. Upper plot correspond to pure Si and lower plot to Si$_{0.8}$Ge$_{0.2}$. The green filled line is the kinetic fraction and the red line the collective fraction of the thermal conductivity. The black line is the total thermal conductivity.}
\label{fig:k_size}
\end{figure} 

Figure~\ref{fig:alloy_bulk} shows the good prediction given by the model for bulk alloys (7 mm wide rods) at 300 K. The black line over the filled area shows the total thermal conductivity, and the green and red areas show the kinetic and collective fraction of the total transport, respectively. Note that collective transport is only important for very pure materials (Si or Ge), with impurity fractions of the order of 0.5\% already destroying most of the collective regime. The prediction given by the current proposal in this region is in remarkable agreement with experimental results from \cite{Cahill2005}, showing that in order to have a model able to predict at the same time the thermal conductivity for pure and alloyed materials, proper description of the evolution of the collective contribution is needed, which is achieved in the KCM by the inclusion of $\Sigma$. This might be the reason why pure kinetic expressions need an extra parameter in the alloy term~\cite{Wang2010b}.

In Fig.~\ref{fig:k_size} we represent the kinetic and collective contributions for two different situations, for pure silicon at the top and Si$_{0.8}$Ge$_{0.2}$ at the bottom. In pure Si, it can be observed that in the region where the thermal conductivity is not affected by size effects, the kinetic/collective fraction does not change significatively. When boundary effects are noticeable, the reduction in the collective flux is more important and the ratio kinetic/collective raises. This is because boundary scattering is resistive and its appearance reduces the weight of the collective term. We notice that the disappearance of the collective contribution is the responsible for a much stronger decrease in thermal conductivity in the $L_{\rm eff}=$1~$\mu$m--10~nm range compared to the micron and above scale. \footnote{Nevertheless, quantum effects should become important below 30~nm ~\cite{DeTomas2014}}. Present results agree with recent works pointing on the direction that long mean free path phonons are important to predict thermal conductivity in alloy samples \cite{Cheaito2012,Vermeersch2015,Vermeersch2015a}. For the alloy sample, in the bottom figure, the collective term has been completely eliminated due to the impurity scattering  and boundary scattering does not change the regime. Also, the functional dependence with sample size is completely different in both samples. The alloy shows the $L^{1/4}$ divergence expected for kinetic samples dominated by boundary scattering~\cite{Wang2010b} while the change in the slope in the pure Si case seems to be behind the destruction of the collective term regime, leaving a linear dependence with size at low temperatures. High-frequency phonons in the collective regime drag low frequency phonons, and the increase with sample size is much reduced at high temperatures~\cite{Tomas}.

\begin{figure}[t!]
\includegraphics[width=\columnwidth]{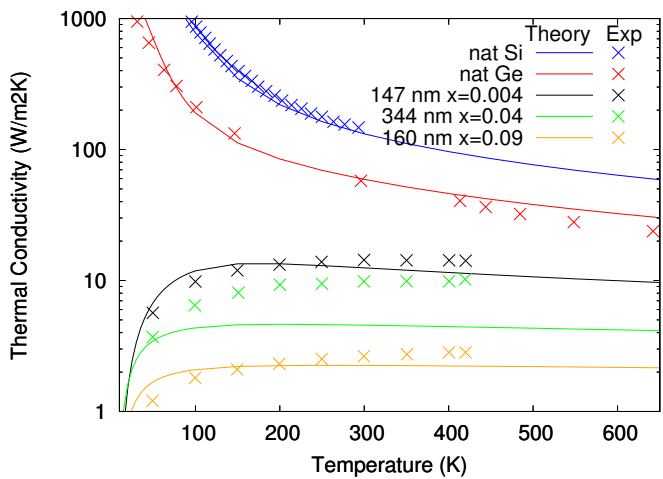}
\caption{Thermal conductivity in terms of Temperature for samples with Ge content and size. Curves are computed according to Eqs.~(\ref{kappa_final})-(\ref{eq:S2}). Experimental data are obtained from \cite{Kim2010}\label{fig:alloy_nw}}
\end{figure}

To show the predictive capability of the KCM in nanometer alloyed samples, in Figure~\ref{fig:alloy_nw} we have plotted the thermal conductivity for bulk silicon and germanium, as well as a selection of Si/Ge alloy nanowires. All the curves, showing a remarkable agreement with the experimental points, are calculated from the same single fit to natural bulk silicon, indicating that the change in the collective/kinetic ratio seems to be the key in predicting this kind of attribute. For pure Si nanowires~\cite{Li2003}, alternative interpretations based on a purely diffusive viewpoint have been proposed\cite{Martin2009b,Chen2008}; nevertheless the KCM has been able to adjust the results by Li {\it et al.}~\cite{DeTomas2014a} as well as the SiGe alloy nanowires in Fig.~\ref{fig:alloy_nw}. While surface roughness in nanowires may have an important impact on thermal conductivity~\cite{Chen2008,Martin2009b}, we haven't included it in our analysis because the wires in Ref.~\onlinecite{Kim2010} had sub-nm roughness.

In conclusion, thermal conductivity values derived from the Kinetic Collective Model for pure Si, Ge and alloy Si/Ge samples ranging from bulk to nanowires for a wide range of temperatures show excellent agreement with experimental data, based on a single fit to natural bulk silicon. No further adjustable parameters are needed, showing that the difficulty in previous attempts to fit wide ranges of temperature, sizes and compositions seems to be related with the transition from the collective to the kinetic regime experienced by a sample with increasing resistive scattering. This, together with the fact that the expressions for the relaxations times agree with theoretical predictions without any modification, gives plausibility to the model. Boundaries and species variation reduce the collective transport in alloys, with alloy scattering a more effective mechanism than boundary scattering. Alloy concentrations as little as 0.5\% or system sizes of the order of 10~nm will destroy the collective contribution to the thermal conductivity. This insight could be very relevant when addressing phenomena such as phonon drag or dopant effects in semiconductor thermoelectrics.

This work has been partially funded by the Spanish Ministerio de Econom\'ia y Competitividad under Project Nos. TEC2012-31330, FIS2012-32099, MAT2012-33483 and Consolider nanoTHERM CSD2010-00044 and Generalitat de Catalunya under Project 2014-SGR0064. Also, the research leading to these results has received funding from the European Union Seventh Framework Programme under grant agreement No. 604391 Graphene Flagship.

\bibliographystyle{apsrev}

\end{document}